\documentclass[prx,reprint,superscriptaddress,showpacs,showkeys,longbibliography]{revtex4-2}
\usepackage{amsmath,amsfonts,amssymb,bm,graphicx,color,multirow,tabularx}

\usepackage[colorlinks=true,linkcolor=blue,urlcolor=blue,citecolor=blue]{hyperref}

\makeatletter
\newcommand*{\balancecolsandclearpage}{%
  \close@column@grid
  \cleardoublepage
  \twocolumngrid
}
\makeatother

\begin{document}

\title{Collapse of Metallicity and High-$T_c$ Superconductivity \\ in the High-Pressure phase of  FeSe$_{0.89}$S$_{0.11}$}

\author{Pascal Reiss}
\email[]{p.reiss@fkf.mpg.de}
\affiliation{Clarendon Laboratory, Department of Physics, University of Oxford, Oxford, UK}
\affiliation{Max Planck Institute for Solid State Research, Stuttgart, Germany}

\author{Alix McCollam}
\affiliation{High Field Magnet Laboratory (HFML-EMFL), Radboud University, 6525 ED Nijmegen, The Netherlands}

\author{Zachary Zajicek}
\affiliation{Clarendon Laboratory, Department of Physics, University of Oxford, Oxford, UK}

\author{Amir A. Haghighirad}
\affiliation{Clarendon Laboratory, Department of Physics, University of Oxford, Oxford, UK}
\affiliation{Institute for Quantum Materials and Technologies, Karlsruhe Institute of Technology, Karlsruhe, Germany}

\author{Amalia I. Coldea}
\email[]{amalia.coldea@physics.ox.ac.uk}
\affiliation{Clarendon Laboratory, Department of Physics, University of Oxford, Oxford, UK}

\date{\today}

\begin{abstract}
We investigate the high-pressure phase of the iron-based superconductor FeSe$_{0.89}$S$_{0.11}$ using transport and tunnel diode oscillator studies. We construct detailed pressure-temperature phase diagrams that indicate that outside of the nematic phase, the superconducting critical temperature reaches a minimum before it is quickly enhanced towards 40\,K above 4\,GPa. The resistivity data reveal signatures of a fan-like structure of non-Fermi liquid behaviour which could indicate the existence of a putative quantum critical point buried underneath the superconducting dome around 4.3\,GPa. Further increasing the pressure, the zero-field electrical resistivity develops a non-metallic temperature dependence and the superconducting transition broadens significantly. Eventually, the system fails to reach a fully zero-resistance state despite a continuous finite superconducting transition temperature, and any remaining resistance at low temperatures becomes strongly current-dependent. Our results suggest that the high-pressure, high-$T_c$ phase of iron chalcogenides is very fragile and sensitive to uniaxial effects of the pressure medium, cell design and sample thickness which can trigger a first-order transition. These high-pressure regions could be understood assuming a real-space phase separation caused by concomitant electronic and structural instabilities. 
\end{abstract}


\maketitle

\footnotetext[1]{See Supplemental Material at [URL will be inserted by publisher] for further experimental data and analyses.}

In the quest of seeking higher and higher superconducting transition temperatures, the application of large hydrostatic pressures is an important tool. While the highest transition temperatures  close to room temperatures were observed in conventional hydride superconductors under enormous pressures in the mega-pascal range \cite{Drozdov2015,Drozdov2019,Grockowiak2022}, much lower pressures in the giga-pascal range are sufficient to boost superconductivity in unconventional superconductors, most notably in the copper-based \cite{Chu1993} and in the iron-based \cite{Sun2012} systems.

The family of the iron-chalcogenide FeSe has emerged as an enormously versatile system in which superconductivity can be tuned not only by hydrostatic pressure, but also by uniaxial strain, isovalent and charge doping, surface dosing, and chemical intercalation \cite{Ge2014,Burrard-Lucas2013,Sun2015,Wen2016,Sun2017,Reiss2017,Farrar2020,Ghini2021,Zajicek2022}. The richness of this system partially stems from its instability towards an electronic nematic phase at ambient conditions, and its proximity to a magnetic phase which is stabilized under high pressures \cite{Sun2015}. Remarkably, it is possible to disentangle the electronic nematic and magnetic phases through a careful combination of iso-electronic doping of FeSe$_{1-x}$S$_x$ with hydrostatic pressures \cite{Matsuura2017}, which allows one to study the contributions of their corresponding order parameter fluctuations to the superconducting pairing independently. 

In particular, applied pressure studies of FeSe$_{0.89}$S$_{0.11}$ have provided unique access to an isolated nematic quantum critical point \cite{Reiss2019,Reiss2020b}. Towards higher pressures, multiple studies identified a strongly enhanced superconducting phase with the transition temperature $T_c$ approaching values of $\approx 35$\,K. However, the nature of this high-pressure phase and of the underlying electronic structure remains unclear, and previous studies produced partially contradictory results. In particular, a resistivity study reported enhanced superconductivity up to 8\,GPa, and the emergence of a seemingly competing spin-density magnetic phase in the pressure range $2-4$\,GPa \cite{Matsuura2017}. In contrast, a subsequent $ac$ susceptibility study did not find signatures of any magnetic order but observed a complete loss of the superconducting shielding at $4$\,GPa \cite{Yip2017}. Another recent transport and NMR study detected weak signatures for magnetism above $3$\,GPa, and a gradual suppression of superconductivity beyond $4$\,GPa \cite{Rana2020}. In Cu$_x$Fe$_{1-x}$Se only magnetism was suppressed under pressure but superconductivity remained strong \cite{Zajicek2022b}, whereas in thin flakes of FeSe, an unsual suppression of the magnetic and superconducting phases under pressure with decreasing flake thickness was observed \cite{Xie2021}. Similarly, the sample thickness and the choice of pressure medium were found to strongly affect the boundaries of the superconducting phase in bulk FeSe above $5$\,GPa, likely due to additional uniaxial stress along the $c$ direction \cite{Miyoshi2021}.

In order to assess the nature of this enigmatic high-pressure phase in detail, we investigate the high-pressure regime of FeSe$_{0.89}$S$_{0.11}$ using transport and tunnel-diode oscillator (TDO) probes inside opposing Diamond Anvil Cells (DACs) up to $\approx 7$\,GPa. We find that the superconducting transition temperature $T_c$ shows a minimum around $ \approx 1$\,GPa, before it increases towards $40$\,K (onset) above $\approx 4$\,GPa. Above $p_s \approx 4$\,GPa, we observe significant changes in the samples properties as the room-temperature normal state resistivity displays a sudden increase with pressure, whereas the low-temperature resistivity develops a marked fan of non-Fermi liquid behavior centered at $p_s$. The superconducting transition broadens significantly above $p_s$ until it becomes incomplete as it fails to induce a zero-resistance state even at the lowest temperatures, where we detect an unusually small critical current density.
Our study suggests that the high-pressure phase of single-crystalline FeSe$_{0.89}$S$_{0.11}$ suffers a structural instability above $p_s$, likely to an orthorhombic symmetry, as reported on FeSe \cite{Sun2015,Svitlyk2017}. As a result of the first-order nature of the structural transition, the system shows a pressure-induced cross-over from a metallic-like to a insulating-like behaviour which harbours a phase coexistence between superconducting/metallic and non-superconducting/semi-metallic domains. Within this picture, the observation of a quantum critical fan as a typical signature of a second-order electronic instability is unusual and it is interrupted by the first-order transition.

\begin{figure*}[t!]
	\includegraphics[trim={0cm 0cm 0cm 0cm}, width=1\linewidth,clip=true]{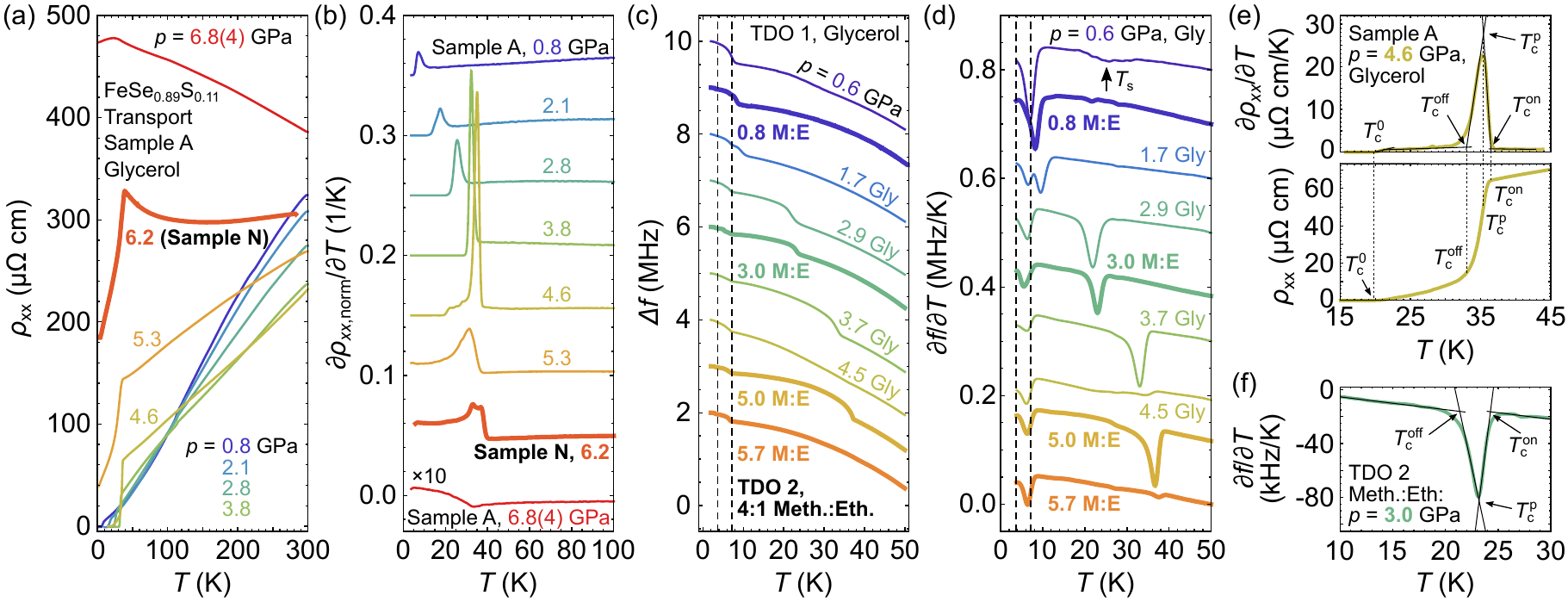}
	\caption{Evolution of the zero-field properties of FeSe$_{0.89}$S$_{0.11}$.
	(a) The temperature and pressure dependencies of the zero-field resistivities of samples A (thin lines) and sample N (thick line). The pressure transmitting medium for both samples is Glycerol.
	(b) First derivative of the low-temperature regime of panel (a) focusing on the superconducting transitions. Here, the resistivities are normalized against their values at $T = 100$\,K and their derivatives shifted vertically for clarity.
	(c) Low-temperature evolution of the resonant frequency of a TDO circuit comparing the effects of either Glycerol as pressure transmitting medium (TDO 1, `Gly', thin lines), or a 4:1 mixture of Methanol:Ethanol (TDO 2, `M:E', thick lines).
	(d) First derivative of the TDO resonant frequency. $T_s$ indicates the position of the nematic transition visible at lowest pressures. Vertical dashed lines in panels (c) and (d) indicate the pressure-independent superconducting transitions of In, Sn and Pb, which originate from the solder joints outside the pressure cell.
	(e,f) Definition of the superconducting critical temperatures indicated by arrows. The onset and offset temperatures, $T_c^\text{on}$ and $T_c^\text{off}$, are defined where the derivative of the resistivity or the TDO resonant frequency recovers their high- and low-temperature values, respectively. The peak in the derivative defines $T_c^\text{p}$, and $T_c^0$ is the temperature with truly zero resistance. At high pressures $p \gtrsim 4$\,GPa, an additional low-temperature tail in resistivity leads to $T_c^0 \ll T_c^\text{off}$.
}
	\label{fig:data}
\end{figure*}

\section{Methods}
\label{sec:methods}
Single crystals of FeSe$_{1-x}$S$_x$ with $x = 0.11$ sulfur substitution were grown using the KCl/AlCl$_3$ chemical vapor transport method as described elsewhere \cite{Bohmer2016}. High-pressure measurements were carried out using an opposing Diamond Anvil Cell (DAC) with a design similar to Refs.~\citenum{Moulding2020} and \citenum{Moulding2022}, using $800$\,$\mu$m bevelled culets. BeCu gaskets with an initial thickness between $400$-$450$\,$\mu$m were pre-indented to $\approx 80$\,$\mu$m, and were subsequently drilled to produce a pressure chamber with a diameter of $400$\,$\mu$m. A thin layer of Stycast 1266:Alumina mixture was applied as gasket insulation, and the gasket was drilled again with a diameter of $350$\,$\mu$m. A total of five single crystals, all roughly $150 \times 70 \times 30~\mu \text{m}^3$, in dimensions were subjected to high-pressure transport measurements inside the DAC using Glycerol as pressure medium. Samples A and B were measured using a standard 4 contact configuration with the voltage contacts placed onto one of the longest sample edges to determine the longitudinal resistivity. For samples C and D, the voltage contacts were placed on opposing edges of the sample suitable for a Hall effect measurement. Sample N was measured using a 5 contact setup, allowing simultaneous longitudinal and Hall effect measurements. For comparison in the low-pressure limit, sample E was measured under ambient pressure only, equally with a 5 contact layout. Sample P was measured using a 5 contact layout inside a piston cylinder cell (PCC) using Daphne Oil 7373 as pressure transmitting medium, as previously reported in Refs.~\citenum{Reiss2019} and \citenum{Reiss2020b} (called Sample A there). All transport measurements were carried out using the AC LockIn technique with a low frequency and a low excitation current $I_p = 1$\,mA within the $(ab)$ plane, unless stated otherwise. A further two single crystals were studied using the Tunnel Diode Oscillator (TDO) technique, using either Glycerol (TDO1) or a 4:1 mixture of Methanol:Ethanol (TDO2) as pressure media. The latter ensures hydrostatic conditions up to $10$\,GPa, well beyond the highest pressure reported here, whereas the former shows non-hydrostatic behavior from 4-6\,GPa onwards \cite{Jayaraman1983,Tateiwa2009,Moulding2020,Moulding2022}. In the TDO studies, we identify pressure-independent signatures associated with the superconducting transition temperatures of Pb, Sn and In, which occur in the solder joints outside the pressure cell. When they become superconducting, the quality factor of the $LC$ circuit changes which affects the resonant frequency $f$. For the DACs, we used the Ruby fluorescence technique at room temperature to determine the pressure inside the cell using multiple small Ruby chips. The reported pressures here correspond to the average before and after cooldown, and error bars indicate the difference which was usually below $0.1$\,GPa. For the piston cylinder cell used for sample P, the pressure was determined from the superconducting transition temperature of Sn at low-temperatures. In order to compare the low-$T$ pressure scale of sample P reported before \cite{Reiss2019,Reiss2020b} with the room-temperature pressure scales of the other samples reported here, the pressures determined for sample P were shifted by $+0.2$ GPa for $p < 1$\,GPa and $+0.1$ GPa for $p < 2$\,GPa to account for the pressure losses of Daphne 7373 during cool-downs. All high-pressure measurements were carried out on compression. Pressures were changed at room-temperature and the cells were allowed to relax until no change in pressure and the applied load could be resolved. Cooling rates were $0.5$\,K/min except for sample N where higher rates were used.

For a quantitative assessment of the evolution of superconductivity, we define four different superconducting transition temperatures, as exemplified in Figs.~\ref{fig:data}(e) and (f). Firstly, by fitting the derivatives of the resistivity and the TDO resonant frequency, $\partial \rho / \partial T$ or $\partial f / \partial T$, respectively, the onset temperature $T_c^\text{on}$ is extracted from the crossing of tangents fitted to the high-temperature normal state data and the leading edge of the superconducting transition. Secondly, the peak temperature $T_c^\text{p}$ where the first derivatives show a maximum or minimum, respectively, is extracted from the crossing of tangents fitted to the leading and trailing edges of the transition. Thirdly, the offset temperature $T_c^\text{off}$ is defined analogously to $T_c^\text{on}$, but using the trailing edge and the low-temperature data. Fourthly, we extract the true zero-resistance temperature $T_c^0$ when the signal falls below the noise level.

\section{Normal state and superconducting properties in zero magnetic field}
\label{sec:zeroField}

\subsection{Phase Diagram}

Previous studies have established that at ambient pressure FeSe$_{0.89}$S$_{0.11}$ becomes superconducting below a sharp transition temperature $T_c \approx 10.4$\,K inside a nematic electronic phase below $T_s \approx 60$\,K \cite{Reiss2017,Xiang2017,Reiss2019,Coldea2016,Reiss2020b,Rana2020}. Upon application of hydrostatic pressure, the nematic transition temperature is quickly suppressed and a nematic quantum critical point is revealed around $p_c \approx 0.6$\,GPa as determined at low temperatures (corresponding to approximately $0.8$\,GPa at the room-temperature pressure). Quantum oscillations have revealed the presence of a Lifshitz-like transition across the border of the nematic phase while the effective masses of the quasiparticles do not display divergent behaviour \cite{Coldea2016,Reiss2019}.

Figures~\ref{fig:data}(a-d) show the temperature dependence of the zero-field resistivity and TDO resonant frequency for different single crystals of FeSe$_{0.89}$S$_{0.11}$ up to $p \approx 7$\,GPa. At the lowest pressure accessible with the Diamond Anvil Cell, the TDO data for $p = 0.6(1)$\,GPa shows a weak anomaly around $25$\,K,  best visible as a dip in the temperature derivative shown in Fig.~\ref{fig:data}(d). We associate this anomaly with the nematic phase transition, in good agreement with previous reports using piston cylinder cells \cite{Reiss2019}. Moreover, Figs.~\ref{fig:data}(a-d) reveal clear signatures of superconducting phase transitions as either a sharp drop in the resistance, or a surge in the TDO resonant frequency. Interestingly, above $4.2$\,GPa the superconducting transitions broaden significantly and become incomplete as the sharp drop in the resistance does not lead to a zero-resistance state anymore, as shown in Fig.~\ref{fig:data}(a). Therefore, in order to quantify the broadening and the loss of a fully superconducting phase, we will follow the evolution of four different superconducting transition temperatures, corresponding to the onset ($T_c^\text{on}$), sharpest drop ($T_c^\text{p}$), offset ($T_c^\text{off}$) and true zero-resistance ($T_c^0$), as indicated in Figs.~\ref{fig:data}(e-f) and the Methods section.

\begin{figure}[t!]
	\includegraphics[trim={0cm 0cm 0cm 0cm}, width=1\linewidth,clip=true]{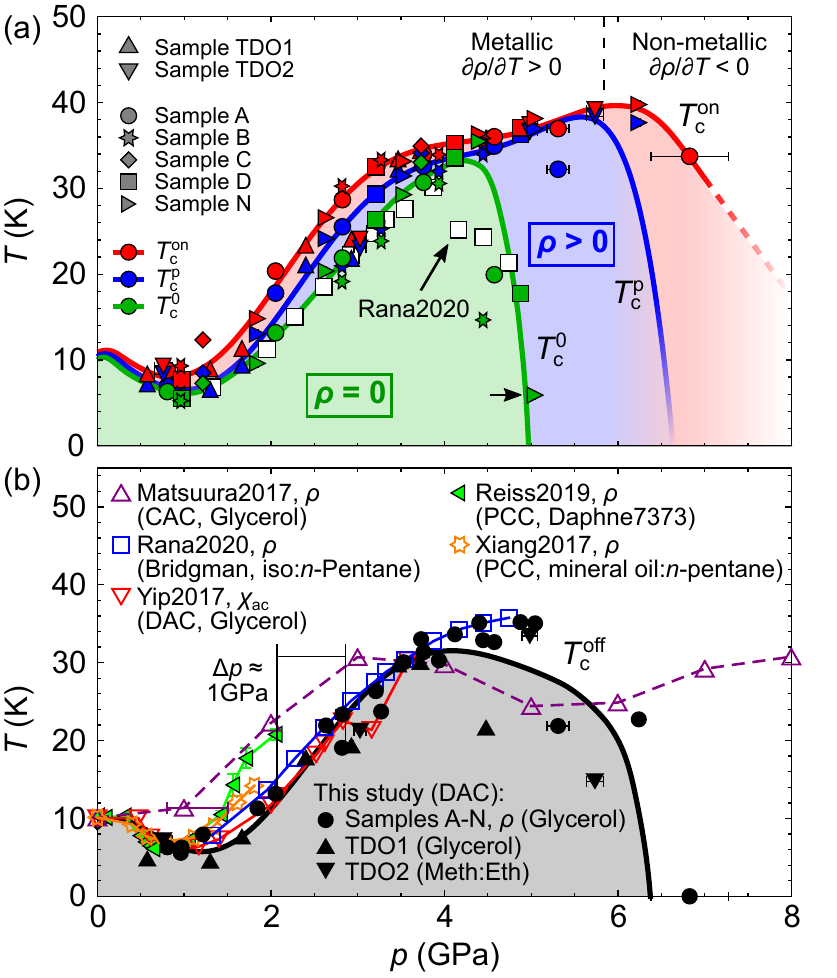}
	\caption{Pressure dependence of the different superconducting transition temperatures $T_c$. 
	(a) The pressure dependencies of the onset temperature $T_c^\text{on}$, the temperature of the steepest change of the resistance or the TDO signal $T_c^\text{p}$, and the zero-resistance temperature $T_c^0$. These transition temperatures start deviating from each other beyond $\approx 4$\,GPa. The open (white) symbols correspond to $T_c^0$ from a previous report for FeSe$_{0.91}$S$_{0.09}$ in Ref.~\cite{Rana2020}. A strong current dependence is observed for the data point marked by an arrow ($T_c^0$ of sample N at $p = 5.2$\,GPa), as discussed in the main text. Thick solid lines are guide-to-the-eyes and the dashed line is a linear extrapolation.
	(b) Comparison of the offset superconducting transition temperatures $T_c^\text{off}$ using different high-pressure methods, as reported here and previously. CAC = Cubic Anvil Cell, DAC = opposing Diamond Anvil Cell, PCC = Piston Cylinder Cell, Meth:Eth = 4:1 mixture of Methanol:Ethanol, $\rho$ = transport study, $\chi_{ac}$ = AC susceptibility study \cite{Yip2017,Matsuura2017,Xiang2017,Reiss2019,Rana2020}.}
	\label{fig:tc}
\end{figure}

Figure~\ref{fig:tc} summarizes the evolution of the different superconducting transition temperatures in a detailed pressure-temperature phase diagram for all single crystals of FeSe$_{0.89}$S$_{0.11}$ studied here and in previous reports \cite{Yip2017,Matsuura2017,Xiang2017,Reiss2019,Rana2020}. At the lowest pressures, the superconducting transition temperatures $T_c$ drop from $\approx 10$\,K at ambient pressure towards a minimum $\approx 6$\,K at a pressure of $\approx 1$\,GPa, independent of the pressure technique or medium employed. In the intermediate pressure regime between $1\,\text{GPa} < p \lesssim 4\,\text{GPa}$, superconductivity is strongly enhanced and all critical temperatures continue to follow each other closely, reaching values close to $35$\,K at 4.0 GPa. However, the highest values for $T_c$ are obtained for pressure techniques which ensure a more hydrostatic environment, such a piston cylinder cells or cubic anvil cells \cite{Matsuura2017,Xiang2017,Reiss2019}. In contrast, studies performed using opposing anvil cells, as in this report and Ref.~\citenum{Yip2017}, display transition temperatures up to $10$\,K lower, or shift the high-pressure, high-$T_c$ phase to higher pressures by $\Delta p \approx 1$\,GPa, as indicated in Fig.~\ref{fig:tc}(b).

In the high pressure regime above $ \approx 4$\,GPa, the critical temperatures display divergent trends. Fig.~\ref{fig:tc} shows that the onset temperature $T_c^\text{on}$ keeps increasing for all samples studied and reaches a maximum close to $40$\,K around $p \approx 6$\,GPa, remarkably similar to the maximum superconducting temperature of $38.3$\,K detected in FeSe at $6.3$\,GPa \cite{Sun2017}. In contrast, the zero-resistance critical temperature $T_c^0$ and the temperature $T_c^\text{p}$, marking the sharpest drop in the resistivity or the surge in the TDO resonant frequency, respectively, become strongly sample dependent. On average, $T_c^0$ drops quickly and vanishes around $p \approx 5$\,GPa, consistent with a previous report \cite{Rana2020}, whereas $T_c^\text{p}$ disappears around $p \approx 6$\,GPa. These distinct pressure dependencies of the critical temperatures $T_c^0$, $T_c^\text{p}$ and $T_c^\text{on}$ suggest that the loss of superconductivity does not imply a closure of the gap, but rather an inhomogeneous and/or filamentary nature of superconductivity. This finding will be corroborated by the critical current studies that indicate a strongly reduced superconducting volume fraction, as discussed in Section~\ref{sec:IV}. Moreover, the alteration of the superconducting phase is consistent with a previous $ac$ susceptibility study which identified a weakened diamagnetic shielding in the same pressure regime \cite{Yip2017}.

\begin{figure}[t!]
	\includegraphics[trim={0cm 0cm 0cm 0cm}, width=1\linewidth,clip=true]{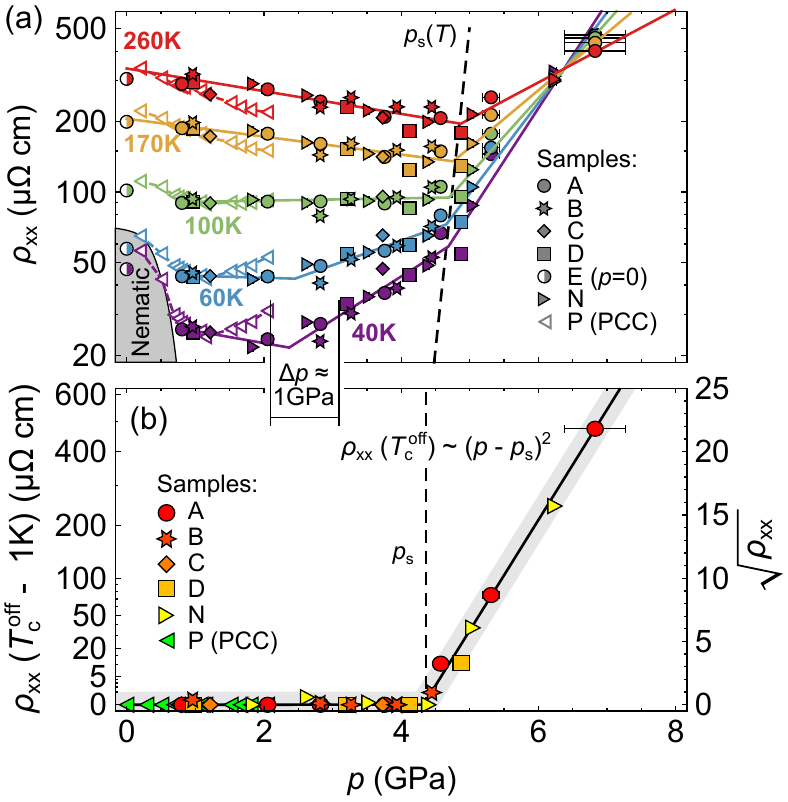}
	\caption{Correlation between resistivity increase and loss of superconductivity. 
	(a) Absolute resistivity as a function of pressure at selected fixed temperatures. The horizontal bar indicates a pressure offset of $\Delta p \approx 1$\,GPa between the data measured on sample P using a piston cylinder cell (open triangles) and the other samples using Diamond Anvil Cells (filled symbols). 
	(b) Residual resistivity just below $T_c^\text{off}$. Solid lines are a guides to the eye. The dashed lines indicate the relevant pressure scale where kinks in the data appear.}
	\label{fig:pressureScale}
\end{figure}

Mirroring the evolution of the superconducting transition temperatures with pressure, the overall behaviour of the resistivity is also very sensitive to the applied pressure in different regimes. To better quantify the relevant pressure ranges, Fig.~\ref{fig:pressureScale}(a) shows the isothermal evolution of the resistivity for all samples studied here. Initially, and at room temperature, the resistivity shows a continuous reduction for pressures $p \lesssim 5$\,GPa which appears consistent with an increasing electronic bandwidth as the orbital overlaps increase. In contrast, the low-temperature resistivity shows a more nuanced pressure dependence. At the lowest pressures, below $ 1$\,GPa, a marked drop of the low-temperature resistivity is observed as the nematic phase is suppressed. This trend can be explained by a growing Fermi surface and a reduced quasiparticle mass, based on our previous high-pressure transport and quantum oscillation studies up to $2.2$\,GPa \cite{Reiss2019}. Additionally, scattering off nematic and/or spin fluctuations as well as nematic domain boundaries within the nematic phase may further add to the low-pressure, low-temperature resistivity, but all effects get suppressed with increasing pressure \cite{Wiecki2018,Kuwayama2019}. Upon further increasing the pressure, the low-temperature resistivity shows an upturn around $\approx 1.2$\,GPa (sample P) or $\approx 2.2$\,GPa (samples A-N). The corresponding pressure difference $\Delta p \approx 1$\,GPa between the resistivities of sample P and samples A-N is consistent with the pressure difference of their superconducting transition temperatures, as discussed above. Interestingly, for an intermediate temperature $T \approx 100$\,K, the resistivity appears essentially pressure independent between $\approx 1$\,GPa and $ \approx 4$\,GPa, which corresponds to the visible crossing point of the resistivity curves in Fig.~\ref{fig:data}(a), similar to previous reports \cite{Reiss2019,Rana2020}. Such a crossing could correspond to an electronic crossover from incoherent to coherent transport predicted for a Hund's metal, or to a change in scattering or electronic correlations tuned by pressure \cite{Haule2009}.

Towards highest pressures, the evolution of the resistivity changes significantly at all temperatures. Figure~\ref{fig:pressureScale}(a) shows an upturn in the resistivity as a function of pressure at $p_s(T) \approx 4.5 - 5$\,GPa. Since $p_s$ can be identified up to room temperature, we interpret it as a signature of a pressure induced structural transition. Figure~\ref{fig:pressureScale}(b) shows the residual resistivity just below the offset temperature $T_c^\text{off}$ which illustrates that the zero-resistivity superconducting state is lost above  $p_s(T_c^\text{off}) \approx 4.2$\,GPa. At higher pressures, the residual resistivity increases sharply as $\propto (p - p_s)^2$, signaling the emergence of a low-temperature resistive tail below the main superconducting transition. Eventually, the resistivity displays incomplete superconducting transition beyond $p \approx 5$\,GPa, as discussed above. This unusual loss of superconductivity is similar to previous studies of FeSe, where it was associated with a structural transition into a orthorhombic or possibly hexagonal phase \cite{Medvedev2009,Sun2015}. Therefore, the evolution of the normal and superconducting behavior at $p = p_s$ appear related and could be driven by a structural transition.

Importantly, the pressure $p_s \approx 4.5$\,GPa, at which there are substantial changes in the resistive and superconducting properties of FeSe$_{0.89}$S$_{0.11}$, is similar to the hydrostatic limit of the employed pressure medium Glycerol \cite{Tateiwa2009}. In order to test the influence of the pressure medium, we have performed a comparative study using the TDO technique with Glyercol and a 4:1 mixture of Methanol:Ethanol, which ensure hydrostatic conditions up to $10$\,GPa \cite{Jayaraman1983,Tateiwa2009}. Figures~\ref{fig:data}(c) and (d) show the evolution of the resonant frequencies and their first derivatives, respectively, for both pressure media as a function of pressure and temperature. Since an estimation of the sample resistivity from the TDO frequency depends on precise knowledge of the experimental set up (sample and coil dimensions and any parasitic capacitance unavoidable in a pressure cell), we focus on the evolution of the superconducting transition only. Figs.~\ref{fig:data}(c) and (d) reveal that superconductivity is lost for both pressure media, albeit at a pressure about $1$\,GPa larger when using Methanol:Ethanol compared to Glycerol. Therefore, the loss in superconductivity and the increase in resistivity are intrinsic to high-pressure FeSe$_{0.89}$S$_{0.11}$,
but the exact pressure is affected by the transmitting medium.

\begin{figure*}[t!]
	\includegraphics[trim={0cm 0cm 0cm 0cm}, width=1\linewidth,clip=true]{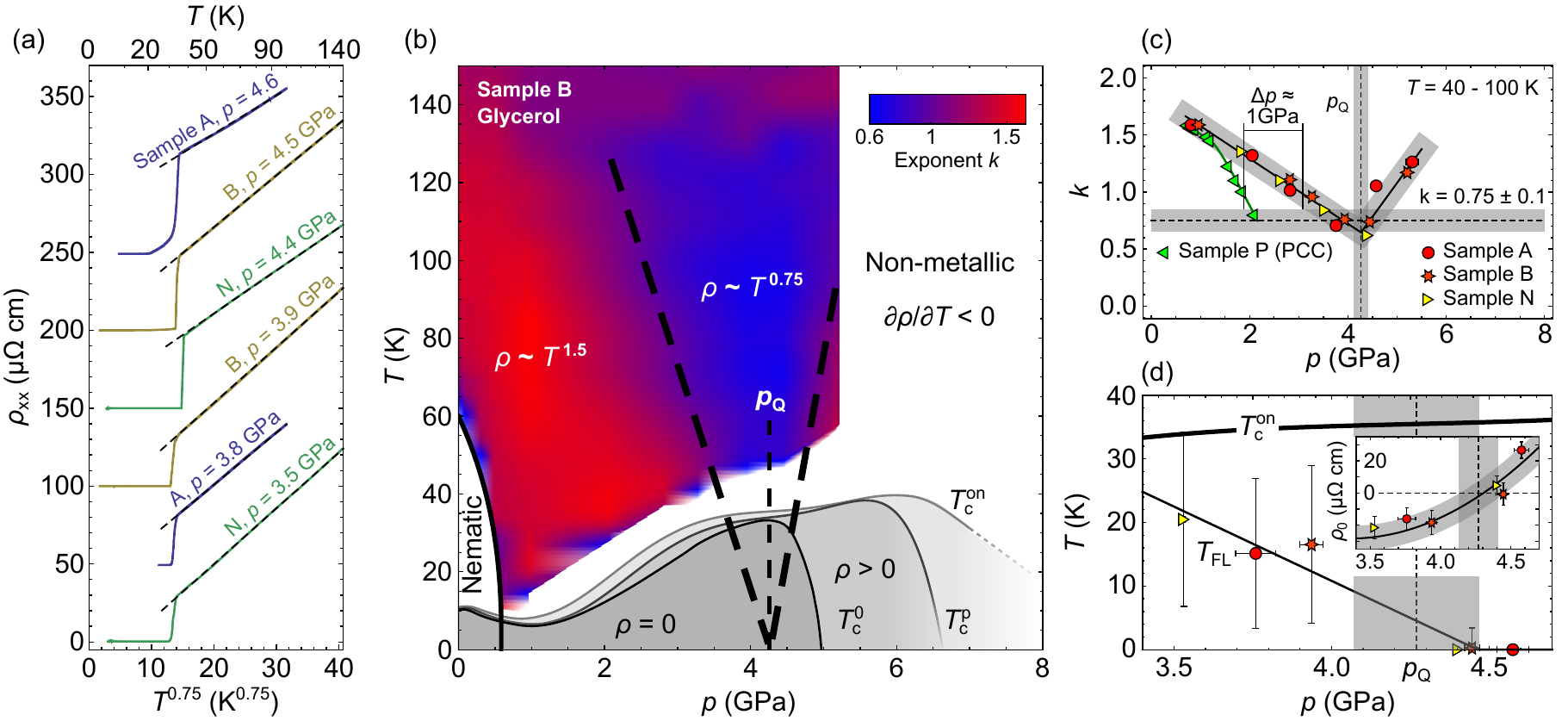}
	\caption{Non-Fermi liquid behavior in pressurized FeSe$_{0.89}$S$_{0.11}$ above the superconducting dome.
	(a) Longitudinal resistivity $\rho_{xx}(T)$ of samples A, B and N plotted against $T^{0.75}$ for pressures close to the maximum of $T_c^0$. Dashed lines are linear fits and curves are vertically offset for clarity.
	(b) Resistivity exponent $k$ as obtained by a the logarithmic derivative of $\partial \rho/\partial T$ of sample B as a function of temperature and pressure. A clear fan with $k \approx 0.75$ is observed above the superconducting dome reaching up to $ \approx 130$\,K, whereas $k \approx 1.5$ outside the fan at low pressures. Dashed lines indicate the boundaries of the fan and their intersection at a pressure $p_Q$ at zero temperature.
	(c) Resistivity exponent $k$ obtained by fitting the experimental data using $\rho_{xx}(T) = \rho_0 + a T^k$ over the fixed temperature interval $ 40 - 100$\,K, outside the nematic phase. The horizontal dashed line indicates the best estimate for $k$ at the minimum and its experimental uncertainty. The vertical dashed line indicates the extrapolated critical pressure $p_Q$ as obtained from panel (b). The finite size horizontal bar indicates a pressure offset $\Delta p \approx 1$\,GPa between the data obtained from sample P measured inside a PCC, and the other samples measured inside DACs.
	(d) The pressure dependence of the Fermi liquid temperature, $T_\text{FL}$, and the superconducting onset temperature $T_c^\text{on}$. The inset shows the residual resistivity $\rho_0$ obtained from fits using $k = 0.75$ which changes sign at $p_Q$. Vertical error bars indicate the standard deviation of $T_\text{FL}$ and $\rho_0$ from varying $k \pm \delta k$ with $\delta k = 0.1$. Solid lines are guides-to-the-eye and the vertical dashed lines represents $p_Q$.}
	\label{fig:NFL}
\end{figure*}

\subsection{Non-Fermi Liquid Behaviour}

Despite the clear changes in the resistive and superconducting properties as well as their mutual correlation as a function of pressure, our multiple studies do not reveal signatures of any additional phase transition. The smooth temperature dependence of the isobar resistivity $\rho(T)$ and the TDO resonant frequency $f(T)$ (see Fig.~\ref{fig:data}), are in marked contrast to previous studies of FeSe$_{1-x}$S$_x$ under high pressures. In the case of FeSe, the resistivity displays kink-like signatures at the onset of a SDW phase from pressure above $\approx 0.8$\,GPa \cite{Sun2015,Terashima2015,Xiang2017,Miyoshi2021}, whereas FeSe$_{0.88}$S$_{0.12}$ shows weaker anomalies in resistivity in the range $ \approx 5-7$\,GPa \cite{Matsuura2017}. We can rule out that any thermodynamic transition involving changes in the magnetic susceptibility occurs even within the superconducting phase as the TDO technique would be capable of detecting such transitions \cite{Lin2020}. Furthermore, superconductivity survives in the high-pressure phase of FeSe and FeSe$_{1-x}$S$_x$ for similar compositions when studied using a cubic anvil cell and no insulating behavior occurs up to $8$\,GPa \cite{Sun2015,Matsuura2017} which clearly differs from our observations. In contrast, the observed loss of superconductivity under pressure observed here is consistent with a previous $ac$ susceptibility study, using a similar pressure technique \cite{Yip2017}. These seemingly contradictory results may point at a strong sensitivity of FeSe-based materials to uniaxial pressure arising from the thermal contraction of opposing anvil cells as employed here \cite{Moulding2020,Moulding2022} and in Ref.~\citenum{Yip2017}, combined with the additional effect induced by the sample thickness \cite{Miyoshi2021,Xie2021}.

Yet, despite the absence of a signature of any long-range phase transition in the temperature dependence of the resistivity and TDO resonant frequency, we observe strong changes in the normal state properties at low temperatures which indicate that our samples are in the proximity to a pressure-induced electronic instability. Figure~\ref{fig:NFL} summarizes the presence of a clear fan of non-Fermi liquid behavior in the resistivity, which is often regarded as a hall-mark of putative quantum critical behavior. The resistivity in the vicinity of $p_s$ displays a power-law form $\rho(T) = \rho_0 + a T^k$ with $k \approx 0.75$ for multiple samples, ranging from the onset of superconductivity up to $ \approx 130$\,K. To trace the pressure and temperature regimes where the exponent $k = 0.75$ provides a valid description of the resistivity, we extract $k$ using a second derivative method as $k = 1 + \partial/\partial(\log T)(\partial \rho / \partial T)$. A key advantage of this method is that it is independent of the knowledge of $\rho_0$ which is often unreliable to extract from high-temperature data. Figure~\ref{fig:NFL}(b) shows the pressure and temperature dependence of the exponent $k$ which reveals a fan-like structure above the superconducting dome for sample B, before the fan is truncated by non-metallic behavior from $ \approx 5$\,GPa onwards. Linearly extrapolating the low- and high-pressure edges of the fan to zero temperature, we find the fan to collapse to a point at $p_Q = 4.3(2)$\,GPa, close to the pressure scale $p_s$ identified above which is also the pressure close to the maximum of $T_c^0$. The presence of a fan-like dependence of the resistivity exponent has been reported in many systems and it is typically regarded as a strong indicator for the presence of a putative quantum critical point (QCP) buried underneath the superconducting dome. A potential candidate for such behaviour could be an antiferromagnetic QCP, suggested to exist in the FeSe$_{1-x}$S$_x$ under pressure, measured using a cubic anvil cell \cite{Sun2015,Matsuura2017}.

Figure~\ref{fig:NFL}(c) shows that the analysis for the resistivity exponent $k$ gives consistent result for all samples measured using DACs, but differs in the pressure dependence from sample P measured in the piston cylinder cell. Here, $k$ was extracted over a fixed temperature interval between $40$\,K and $100$\,K outside the quantum critical fan from low pressures $\lesssim 2$\,GPa and across the quantum critical fan at $p_Q$.  At the lowest pressures, outside the nematic phase, we find $k \approx 1.5$ for the DAC, in excellent agreement with sample P measured inside the piston cylinder cell, consistent with previous reports \cite{Bristow2019,Reiss2019}. For all samples measured in the DACs the exponent $k$ has a sample-independent pressure dependence and it is continuously reduced upon entering the quantum critical fan to reach a minimum of $k = 0.75 \pm 0.1$ for $p_Q = 4.3(2)$\,GPa. Evidently, we detect a pressure offset in the evolution of $k$ in sample P (measured in a piston cylinder cell) by $\Delta p \approx 1$\,GPa around $p \approx 2$\,GPa (Figure~\ref{fig:NFL}(c)), which is consistent with the offset in $T_c$ and the absolute value of $\rho$ in this pressure range, shown in Figs.~\ref{fig:tc}(b) and \ref{fig:pressureScale}(b).

\begin{figure*}[htbp]
	\includegraphics[trim={0cm 0cm 0cm 0cm}, width=1\linewidth,clip=true]{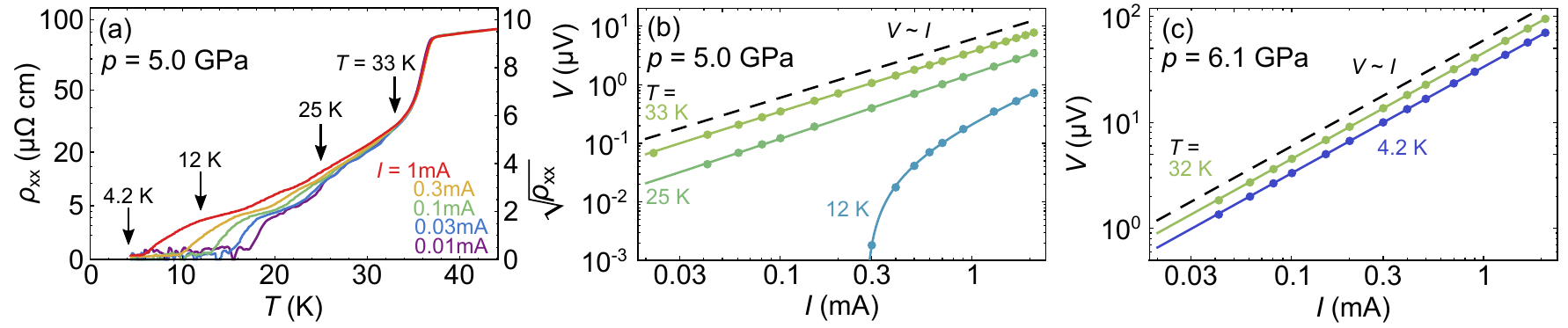}
	\caption{Critical current study in the high-pressure phase of FeSe$_{0.89}$S$_{0.11}$ for sample N.
	(a) The temperature dependence of zero-field resistivity during cooling as a function of applied current $I$ under a pressure of $p \approx 5.0$\,GPa.
	(b-c) $I$-$V$-curves as a function of temperature for $p = 5.0$\,GPa (panel b) and $p = 6.1$\,GPa (panel c). The solid lines are guide-to-the-eyes. The dashed black line is a linear dependence $V \sim I$ indicating an ohmic resistance.
}
	\label{fig:IC}
\end{figure*}

Figure~\ref{fig:NFL}(d) provides an additional test of the location of the quantum critical fan close to $p_Q \approx 4.3$\,GPa. The inset shows the extracted values for $\rho_0$ as obtained from the fixed-exponent fits presented in Fig.~\ref{fig:NFL}(a). Evidently, for small pressures $p < p_Q \approx 4.3$\,GPa, $\rho_0$ appears negative which is unphysical, in contrast to larger pressure $p > p_Q \approx 4.3$\,GPa where a positive $\rho_0$ is obtained. This finding implies that for pressure lower than the putative quantum critical point, the resistivity must return to a larger exponent at low temperatures, e.g. return to Fermi liquid behavior with $k = 2$. We can make a rough estimate for such a Fermi liquid cross-over temperature $T_\text{FL}$ by matching Fermi liquid and non-Fermi liquid behavior and requiring $\rho_0 = 0$. The resulting $T_\text{FL} \approx 10-20$\,K is shown in the main panel of Fig.~\ref{fig:NFL}(d). These values for $T_\text{FL}$ are similar to our previous estimate close to the nematic quantum critical point where $T_\text{FL} = 11$\,K was extracted for sample P \cite{Reiss2019}. Moreover, we find that $T_\text{FL}$ is always lower than $T_c^\text{on}$ and thus cannot be resolved in Figure~\ref{fig:NFL}(b). This analysis also shows that a collapsing $T_\text{FL}$ in the vicinity of $p_Q$ is consistent within the experimental resolution.

Finally, we note that the extracted value of the resistivity exponent $k \approx 0.75 = 3/4$ is unusual and differs from typical values for ferromagnetic or antiferromagnetic quantum criticality, in either 2D or 3D, and for clean or dirty systems \cite{Rosch1999,Brando2015}. Hence, we cannot infer the nature of the QCP from this value. However, theoretical predictions for quantum critical exponents in transport measurements actually relate to the temperature dependence of the scattering rate $\tau^{-1}$, and not to the resistivity $\rho$. Since the resistivity $\rho \sim m^\star \tau^{-1} / n$ also depends on the charge carrier concentration $n$ and the effective mass $m^\star$, an explicit temperature dependence of these quantities will alter the observed resistivity exponent $k$. Indeed, FeSe and related systems display a strong temperature dependence of the chemical potential \cite{Rhodes2017} which manifests as a variation of the charge carrier density as a function of temperature. Additionally, inside the nematic phase, anisotropic scattering due to spin fluctuations may also become important \cite{Farrar2022,Watson2015b}. Further magnetotransport studies will be required to assess changes in scattering in the high-pressure phase of FeSe$_{1-x}$S$_x$.

\section{$I$-$V$ Characteristic}
\label{sec:IV}

We now turn to the nature of the high-pressure, low-temperature phase which displays a finite but incomplete superconducting phase. In order to separate normal metallic from partially superconducting resistances, we perform a $I$-$V$-study to search for non-ohmic conduction and to determine the superconducting critical current density. Assuming an inhomogeneous superconducting phase with only filamentary superconducting paths, we would expect a strongly reduced critical current density $j$, when measured across the entire sample, which should be the smallest close to $T_c$. Figure~\ref{fig:IC}(a) shows cooling curves on sample N at a pressure of $p \approx 5.0$\,GPa under different applied currents across the broad superconducting transition. At this pressure, the sample is tuned beyond $p_s \approx p_Q$, but it shows a distinct superconducting onset, and eventually a zero-resistance for all currents studied, see also Fig.~\ref{fig:tc}. However, in contrast to expectations, we find a strongly current-dependent sample resistivity only for temperatures below $T \approx 20$\,K, whereas close to the onset $T_c^\text{on}$ and sharpest-drop $T_c^\text{p}$, the resistivity is essentially current-independent, as shown in Fig.~\ref{fig:IC}(b). By increasing the pressure towards $6.2$\,GPa, shown in Fig.~\ref{fig:IC}(c), we observe ohmic resistance at any temperature, despite a continuously well defined superconducting onset and sharp drop in the zero-field resistivity, cf. Fig.~\ref{fig:data}(a). This demonstrates that the high-pressure, low-temperature phase beyond $p_s \approx p_Q$ cannot simply be understood as filamentary superconductivity, but a more complex model is required, as discussed below.

\section{Summary and Discussion}
\label{sec:discussion}

\begin{figure}[t!]
	\includegraphics[trim={0cm 0cm 0cm 0cm}, width=1\linewidth,clip=true]{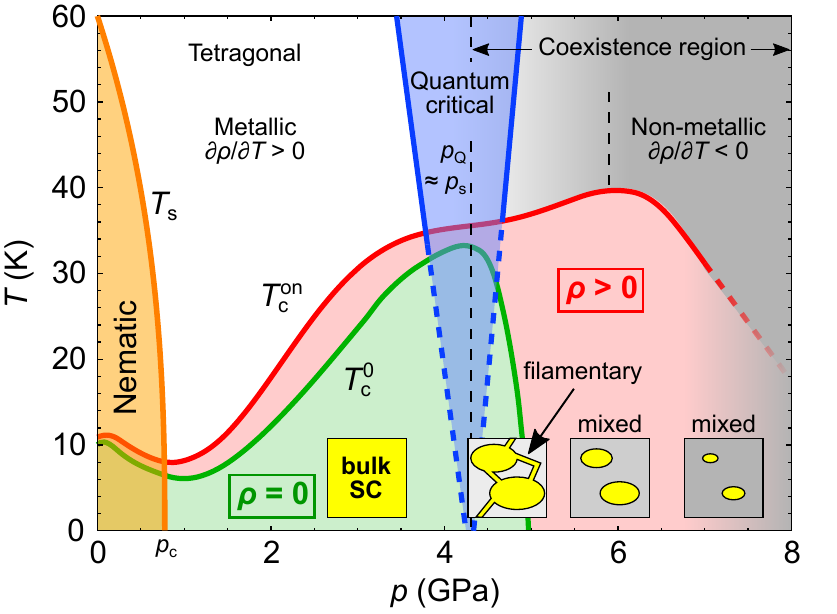}
	\caption{Proposed phase diagram covering the nematic, superconducting, tetragonal, quantum critical and coexistence regions. Real space sketches indicate how the superconducting volume fraction is lost.}
	\label{fig:phaseDiag}
\end{figure}

Our high-pressure study of the superconducting and normal state properties of FeSe$_{0.89}$S$_{0.11}$ reveals a complex pressure-temperature phase diagram, as summarized in Fig.~\ref{fig:phaseDiag}. At very low pressures, the system displays a well-established nematic order below $T_s$ and $p_c$ \cite{Reiss2019,Reiss2020b}, before it
becomes superconducting below $10$\,K. Outside the nematic phase, $T_c$ first reaches a minimum of $6$\,K at a pressure of $1$\,GPa, before it surges towards $\approx 40$\,K with increasing pressure. At high pressures, we observe a number of signatures in our transport data which suggest concomitant electronic and structural instabilities across $p_Q = 4.3$\,GPa and $p_s(T) \approx 4-4.5$\,GPa, respectively. Firstly, reaching up to room temperature, the resistivity shows a marked up-turn at $p_s$, and the system develops a non-metallic temperature dependence. Secondly, we uncover clear signatures of non-Fermi liquid behavior at $p_Q$ which suggests the presence of an electronic quantum critical point located at this pressure. Thirdly, the unusual $I$-$V$ dependence clearly suggests a loss of the superconducting volume fraction, consistent with a previous $ac$ susceptibility study \cite{Yip2017}.

The signatures of the high pressure phase could be explained considering a phase separation scenario, as sketched in Fig.~\ref{fig:phaseDiag}. At low pressures, $p < p_s$, the superconducting phase has sharp superconducting transitions and reaches the zero-resistance state, as shown in Fig.~\ref{fig:data}(a). Moreover, there is a large change in the TDO resonant frequency, as evident from Fig.~\ref{fig:data}(c), and a sizable response in a previous $ac$ susceptibility study \cite{Yip2017}. Therefore, this low-pressure superconducting phase reflects the bulk nature of the superconducting state of the single crystals. In contrast, in the high-pressure phase, a finite resistance remains at lowest temperatures which displays clear non-ohmic behaviour only close to $p_s \approx p_Q$, whereas the onset of the superconducting transition remains sharp and the finite resistance displays ohmic behaviour. This suggests that for temperatures just below $T_c^\text{on}$, only isolated superconducting islands develop within an emerging non-metallic matrix but they fail to connect to provide a fully superconducting current path. Thus, a drop in the resistance is observed when the islands become superconducting, and any finite resistance measured in this temperature regime arises solely from the normal-state non-metallic matrix. In this case, the resistance should have ohmic behaviour and $T_c^\text{on}$ should essentially be current independent, as observed in Fig.~\ref{fig:IC}. Upon further lowering the temperature, the remaining normal-state matrix gradually develops filamentary connections between the superconducting islands, which induce a strongly current dependent phase and non-ohmic resistance, as observed in Fig.~\ref{fig:IC}. With increasing pressure, the superconducting islands shrink and gradually disappear which makes a filamentary connection unlikely. This development is fully consistent with the remaining but shrinking sharp drop of the sample resistance at $T_c^\text{on}$ and $T_c^\text{p}$, even under a pressure of $p = 6.2$\,GPa, the fully ohmic resistance observed at any temperatures, and the lack of a zero-resistance phase. By increasing the pressure even further, the islands are lost eventually, and the drop in resistance disappears, as shown in Figs.~\ref{fig:data}(a).

The occurrence of a phase separation across $p_s \approx p_Q$ is consistent with a structural first-order transition. Yet, the observation of signatures associated with quantum criticality at $p_Q$ implies a second-order transition. The coexistence of such distinct transitions is highly unusual. Previous studies have reported pressure-induced structural transitions in FeSe$_{1-x}$S$_x$ crystals at room temperature, however, they occur at much higher pressures than the low-temperature electronic instabilities. For example, FeSe displays signatures of a low-temperature SDW phase in the pressure range $\approx 2-6$\,GPa which occurs together with a weak orthorhombic lattice distortion associated with the onset of the magnetic order. Yet, the system remains metallic in this pressure regime \cite{Sun2015,Kothapalli2016,Svitlyk2017}. In contrast, a room-temperature structural instability from tetragonal towards orthorhombic and hexagonal phases in FeSe could be induced only at much higher pressures above $10$\,GPa where non-metallic behavior was observed \cite{Sun2015,Medvedev2009,Svitlyk2017}. Similarly for FeSe$_{0.88}$S$_{0.12}$, it was suggested that a SDW phase is stabilized in the pressure range $ \approx 4-7$\,GPa from transport data anomalies, but no signatures of a non-metallic phase emerges at high pressures when using a cubic anvil cell \cite{Matsuura2017}. A pressure-induced structural instability was detected for a similar compound at room temperatures beyond $9$\,GPa only \cite{Nikiforova2020}.

One interesting aspect of the pressure studies of FeSe$_{1-x}$S$_x$ is that any discrepancies could be related to the different pressure media and pressure cell designs used in the experiments as well as the sample thickness. In the case of structural studies, often Helium is used as pressure medium which ensures a much more hydrostatic environment, as compared with Glycerol and 4:1 Methanol:Ethanol used here \cite{Svitlyk2017,Nikiforova2020}. Cubic anvil cells are also less prone to uniaxial thermal contractions than the opposing anvil cell used here \cite{Moulding2020,Moulding2022}, and the resistivities display a zero-resistivity state beyond $4$\,GPa  \cite{Sun2015,Matsuura2017}. In contrast, studied performed with opposing anvil cell designs revealed a gradual loss of superconductivity around $ \approx 4$\,GPa and show good agreement with our study \cite{Yip2017,Rana2020}. This suggests that the concomitant structural and electronic instabilities could arise from finite uniaxial pressure components unavoidable in an opposing anvil cell setup. This assumption is supported by studies which demonstrated that the sample thickness and the choice of pressure medium can have profound influence on the stability of the SDW phase in FeSe \cite{Miyoshi2014,Xie2021}. Thus, it is conceivable that finite uniaxial pressures along the crystallographic $c$ axis might trigger the structural transition around $p_s \approx 4.2$\,GPa. Indeed, many iron-based superconductors are very sensitive to uniaxial pressure components \cite{Gati2020a} and the electronic structure changes significantly with varying chalcogen height above the conducting Fe planes $h=z \cdot c$ \cite{Moon2010}. Interestingly, $p_s \approx p_Q$ is remarkably similar to the hydrostatic limit of Glycerol which solidifies around $4$\,GPa at room temperatures \cite{Tateiwa2009}. However, our TDO study demonstrates that superconductivity survives only to marginally higher pressures of $\approx 5$\,GPa using a 4:1 mixture of Methanol:Ethanol as pressure medium, which has a solidification around $10$\,GPa \cite{Jayaraman1983,Tateiwa2009}. This suggests that the structural and electronic instabilities are not directly triggered by the solidification of Glycerol.

A consequence of the concomitant first-order structural transition at $p_s$ and the second order electronic transition of possibly magnetic origin around $p_Q$ is the inherent domain formation that would naturally lead to additional scattering of the charger-carriers. Interestingly, the resistivity trends observed under pressures can be compared with the evolution of the resistivity in the presence of strong impurity scattering in Cu$_x$Fe$_{1-x}$Se \cite{Gong2021,Zajicek2022}. The primary effect of the Cu-substitution is to disturb significantly the charge carrier transport in the Fe-planes, and thus to lead to a significant enhancement of scattering. As a result, the resistivity shows similar enhancements both in temperature dependence and in absolute values, as compared with those in the high-pressure phase, and additionally there is a strong reduction of charge carrier mobilities  \cite{Zajicek2022,Zajicek2022b}. On the other hand, the loss of superconductivity appears different in Cu$_x$Fe$_{1-x}$Se as the superconducting transition is strongly reduced with increasing Cu content. However, these distinct observations may be reconciled assuming the aforementioned phase separation between metallic/superconducting and non-metallic domains, respectively.

In summary, we have explored the high-pressure phase of the iron-based superconductor FeSe$_{0.89}$S$_{0.11}$.
Our transport and TDO measurements revealed a strongly enhanced superconducting phase at high pressures, which gradually disappears beyond $\approx 4.2$\,GPa as the system becomes non-metallic. This trend suggests the development of a phase separation triggered by an uniaxial pressure component induced first-order structural transition. At low temperatures, we revealed a fan-like structure of non-Fermi liquid behaviour towering above the superconducting phase which points towards a putative quantum critical point buried below. Our results reveal the occurrence of accidentally concomitant structural (first-order) and electronic (second-order) phase transitions which could explain the strongly reduced critical current density of the superconducting phase. These effects reported here emphasize the sensitivity of high-$T_{c}$ superconductivity of iron-chalcogenides to the structural and electronic degrees of freedom induced by uniaxial and hydrostatic pressures.

\section{Acknowledgments}
We thank Sven Friedemann and Patricia Alireza for their valuable technical support setting up the high-pressure techniques in Oxford.
We further acknowledge Matthew Watson for his previous measurement of sample E.
This work was mainly supported by the EPSRC (EP/I004475/1, EP/I017836/1).
P.R.~and A.A.H. acknowledge financial support of the Oxford Quantum Materials Platform Grant (EP/M020517/1).
Part of this work was supported by HFML-RU and LNCMI-CNRS, members of the European Magnetic Field Laboratory (EMFL) and by EPSRC (UK) via its membership to the EMFL (grant no. EP/N01085X/1).
We acknowledge financial support of Oxford University John Fell Fund.
Z.Z. acknowledges financial support from the EPSRC studentship (EP/N509711/1 and EP/R513295/1).
A.I.C.~acknowledges an EPSRC Career Acceleration Fellowship (EP/I004475/1) and Oxford Centre for Applied Superconductivity.

\end{document}